\documentclass[usenatbib]{mn2e}
\usepackage{graphicx,color}
\definecolor{red}{rgb}{0.75,0.0,0.0}
\definecolor{blk}{rgb}{0.0,0.0,0.0}
\definecolor{blu}{rgb}{0.0,0.0,0.75}
\def\hi{H\,{\sc i}}

\def\hei{He\,{\sc i}}

\def\cii{C\,{\sc ii}}

\def\civ{C\,{\sc iv}}
\def\nv{N\,{\sc v}}
\def\oiv{O\,{\sc iv}}
\def\ovi{O\,{\sc vi}}
\def\mgii{Mg\,{\sc ii}}
\def\siiv{Si\,{\sc iv}}
\def\siii{Si\,{\sc ii}}
\def\siiii{Si\,{\sc iii}}
\def\siv{S\,{\sc iv}}
\def\alii{Al\,{\sc ii}}
\def\aliii{Al\,{\sc iii}}
\def\pv{P\,{\sc v}}
\def\feii{Fe\,{\sc ii}}
\def\neviii{Ne\,{\sc viii}}
\def\mgx{Mg\,{\sc x}}
\def\nh{$n_\mathrm{\scriptscriptstyle H}$}
\def\ne{$n_\mathrm{\scriptscriptstyle e}$}
\def\qh{$Q_\mathrm{\scriptscriptstyle H}$}
\def\uh{$U_\mathrm{\scriptscriptstyle H}$}
\def\Nh{$N_\mathrm{\scriptscriptstyle H}$}
\def\mdot{$\dot{M}$}
\def\edot{$\dot{E}_\mathrm{k}$}
\def\lbol{$L_\mathrm{Bol}$}
\def\ledd{$L_\mathrm{Edd}$}
\def\Zsun{\ifmmode {\rm Z}_{\odot} \else Z$_{\odot}$\fi}
\def\kms{\ifmmode {\rm km~s}^{-1} \else km~s$^{-1}$\fi}
\def\Lya{\ifmmode {\rm Ly}\alpha \else Ly$\alpha$\fi}
\newcommand{\revise}[1]{#1}
\newcommand{\secref}[1]{Section~\ref{#1}}
\newcommand{\figref}[1]{Figure~\ref{#1}}
\newcommand{\tabref}[1]{Table~\ref{#1}}
\newcommand{\eqnref}[1]{Equation~(\ref{#1})}
\newcommand{\LL}[1]{\color{blu}$>$&\color{blu}#1&}
\newcommand{\UL}[1]{\color{red}$<$&\color{red}#1&}

\newcommand{\Meas}[3]{&#1&$^{+#2}_{-#3}$}

\def\gtorder{\mathrel{\raise.3ex\hbox{$>$}\mkern-14mu\lower0.6ex\hbox{$\sim$}}}
\def\ltorder{\mathrel{\raise.3ex\hbox{$<$}\mkern-14mu\lower0.6ex\hbox{$\sim$}}}

\title[Energetic quasar outflow of J0831+0354]{Strong Candidate for AGN Feedback: VLT/X-shooter Observations of BALQSO SDSS J0831+0354}
\author[C. Chamberlain et al.]{Carter Chamberlain,$^1$ Nahum Arav,$^1$ Chris Benn$^2$\\
$^1$Department of Physics, Virginia Tech, Blacksburg, VA 24061: carterch@vt.edu\\
$^2$Isaac Newton Group, Apartado 321, 38700 Santa Cruz de La Palma, Spain}
\begin{document}\maketitle

\begin{abstract}
\revise{We measure the location and energetics of a \siv\ BALQSO outflow.}  This ouflow has a velocity of \revise{10,800 \kms} and a kinetic luminosity of $10^{45.7} \mathrm{erg~s}^{-1}$, which is \revise{5.2\%} of the Eddington luminosity of the quasar. From collisional excitation models of the observed \siv$/$\siv* absorption troughs, we measure a hydrogen number density of \nh=$10^{4.3}$ $\mathrm{cm}^{-3}$, which allows us to determine that the outflow is located \revise{110 pc} from the quasar. Since \siv\ is formed in the same ionization phase as \civ, our results can be generalized to the ubiquitous \civ\ BALs. Our accumulated distance measurements suggest that observed BAL outflows are located much farther away from the central source than is generally assumed (0.01-0.1 pc).
\end{abstract}


\section{INTRODUCTION}\label{secIntro}
%Active galactic nuclei (AGN) can interact with their host galaxies through either quasar-mode or radio-mode feedback. Quasar-mode outflows are detected as broad absorption line (BAL) troughs that are blueshifted in the rest-frame spectrum of $\sim$20\% of quasars \citep{Hewett03,Ganguly08,Knigge08}.\color{red}The ubiquity and wide opening angle deduced from the detection rate of these mass outflows allow for more efficient interaction with the surrounding medium\color{black} compared to 

Broad absorption line (BAL) outflows are detected as absorption troughs that are blueshifted in the rest-frame spectrum of 20--40\% of quasars \citep{Hewett03,Ganguly08,Knigge08,Dai08}. 
From their detection rate, we deduce that these outflows cover on average $\sim$20--40\% of the solid angle around the quasar.
Such large opening angles  allow for efficient interaction with the surrounding medium. As shown by simulations, the  mass, momentum and especially the energy carried by these outflows
can  play an important role in the  evolution of galaxies and their environments \citep[e.g.][]{Scannapieco04,Levine05,Hopkins06,Cattaneo09,Ciotti09,Ciotti10,Ostriker10,Gilkis12,Choi14}. 
Theoretical studies show that such interactions can provide an explanation for a variety of observations: the self-regulation of the growth of the supermassive black hole and of the galactic bulge, curtailing the size of massive galaxies, 
and the chemical enrichment of the intergalactic medium. These processes are part of the so-called AGN feedback
 \citep[e.g.][and references therein]{Silk98,dimatteo05,Germain09,Hopkins09,Elvis06,Zubovas14}.

\begin{table}
\caption{Physical properties of energetic quasar outflows.}
\label{tabEnerCompare2}
\begin{tabular*}{\columnwidth}{@{\extracolsep{\fill}}l@{\hspace{0.3cm}}c@{\hspace{0.3cm}}c@{\hspace{0.1cm}}c@{\hspace{0.2cm}}c@{\hspace{0.1cm}}c}\hline
			&v				&$R$				&$\dot{M}$			&Log $\dot{E}_k$	&$\dot{E}_k/L_\mathrm{Edd}^d$\\
Object			&(km/s)				&(pc)				&(M$_\odot$/yr)			&(erg/s)		&(\%)\\\hline
SDSS J0831+0354$^1$	&-10800				&$110^a$			&$135$				&$45.7$			&\revise{5.2}\\	
SDSS J1106+1939$^2$	&-8250	 			&$320^a$			&$390$  			&$46.0$			&\revise{12}\\
HE 0238-1904$^3$	&-5000				&$3400^b$			&$140$				&$45.0$			&$0.7$\\
\revise{SDSS J0838+2955$^4$}	&\revise{-5000}	 			&\revise{3300$^c$}			&\revise{300}				&\revise{45.4}			&\revise{2.3}\\
\revise{SDSS J0318-0600$^5$}	&\revise{-4200}				&\revise{6000$^c$}			&\revise{120}				&\revise{44.8}			&\revise{0.13}\\\hline
\multicolumn{6}{l}{\parbox[t]{\columnwidth}{$R$ from: $^a$high-ionization \siv*$/$\siv; $^b$high-ionization \oiv*$/$\oiv; $^c$low-ionization \siii*$/$\siii}}\\
\multicolumn{6}{l}{\parbox[t]{\columnwidth}{$^d$See \secref{secResultsEner} for $L_\mathrm{Edd}$ determination.}}\\
\multicolumn{6}{l}{\parbox[t]{\columnwidth}{{\bf References.} (1) this work; (2) \citealt{Borguet13}; (3) \citealt{Arav13}; (4) \revise{\citealt{Moe09}}; (5) \revise{\citealt{Dunn10a}}.}}
\end{tabular*}\vspace{-0.3cm}
\end{table}

The importance of quasar outflows to AGN feedback depends on the mass-flow rate (\mdot) and kinetic luminosity (\edot) of the outflowing material. An \edot\  value of at least 0.5\% \citep{Hopkins10} or 5\% \citep{Scannapieco04} of the Eddington luminosity (\ledd) is deemed sufficient to produce the aforementioned feedback effects. A crucial parameter needed to determine \edot\ is the distance $R$ to the outflow from the central source. Lacking spatial image information, we deduce $R$ from the value of the ionization parameter (\uh, see \secref{secPhoto}) once the hydrogen number density (\nh) of the gas is known. Our group has determined \nh\ (leading to $R$, \mdot\ and \edot) for several quasar outflows \citep[e.g.][]{Moe09,Dunn10a,Bautista10,Aoki11,Borguet12a} by utilizing absorption lines from excited states of singly-ionized species (e.g. \feii* and \siii*).  

Absorption troughs from singly-ionized species classify an outflow as a Lo\-BAL. The lower detection rate of Lo\-BALQSO in spectroscopic surveys (3--7\%) compared with 20--40\% for \civ\ BALQSO \citep{Dai12} raises the question of whether the determinations obtained for these objects are representative of the ubiquitous high-ionization \civ\ BALQSO \citep[see][]{Dunn10a}. The most straightforward way to avoid such uncertainty is to observe outflows that show absorption lines from excited states of ions with a similar ionization potential to that of \civ. The optimal ion for ground-based observations is \siv\ \citep[see discussion in ][]{Dunn12,Arav13}.

%The disparate ionization potential between \civ\ ($\mathrm{IP}=47.89$eV) and singly-ionized species ($\mathrm{IP}\sim8$eV) causes the latter to form close to a hydrogen ionization front, while the former (\civ) forms after the first helium ionization front. The physical conditions of the gas in these two spatial regions can differ drastically, therefore the number density deduced from singly-ionized species may not be applicable to the \civ\ region which comprises the bulk of the outflow. A better approach would be to use excited states from ions that form co-spatially with \civ; ions such as \siv\ ($\mathrm{IP}=34.79$eV).

To realize this, we conducted a survey using VLT/X-shooter between 2012 and 2014 aimed at finding quasar outflows showing absorption from \revise{\siv\ $\lambda$1062.66 and the excited state \siv* $\lambda$1072.97}. Of the 24 objects observed, two objects (SDSS J1106+1939 and SDSS J1512+1119) have been published \citep[][hereafter Paper I]{Borguet13}. SDSS J1106+1939 yielded the most energetic \siv\ BAL outflow to date with \edot=$10^{46.0}~\mathrm{erg~s}^{-1}$ (see \secref{secResultsComparison} for further discussion), whereas the SDSS J1512+1119 outflow has \edot$\ltorder 10^{43.8}~\mathrm{erg~s}^{-1}$. In this paper, we present the analysis of an outflow from SDSS J0831+0354, which exhibits similar properties to that of SDSS J1106+1939. This includes the presence of \pv\ and \siv/\siv* troughs, thereby allowing us to determine the distance and energetics (see \secref{secResultsEner}) of the SDSS J0831+0354 outflow. \tabref{tabEnerCompare2} summarizes the current state of the field by listing the energetics of prominent outflows analyzed by our group.

%\cite{Borguet13} (hereafter Paper I) used this method with \siv/\siv* $\lambda\lambda 1062.66,1072.97$ to determine the energetics of the outflow in SDSS J1106+1939, which yielded the most energetic \siv\ BAL outflow to date with \edot=$10^{46.0} \mathrm{erg~s}^{-1}$. The values determined for that outflow, along with other energetic outflows studied by our group, are summarized in \tabref{tabEnerCompare2}. \siv\ has a similar ionization potential to \civ, and since BAL outflows are detected on the basis of their \civ\ BAL, number density diagnostics from \siv\ is directly linked to the same gas that produces the \civ\ absorption. This allows a more solid comparison to be made between the \siv\ BAL outflow of SDSS J1106+1939 and the canonical \civ\ BAL outflows. The outflow of SDSS J1106+1939 was the first detection of absorption from \siv\ that yielded a high \edot. In this paper, we present the analysis of the SDSS J0831+0354 outflow, which exhibits similar properties to SDSS J1106+1939. This includes the presence of \pv\ and \siv/\siv* troughs, thereby allowing us to determine the distance and energetics (see \tabref{tabEnerCompare2}) of the SDSS J0831+0354 outflow.

The plan of this paper is as follows. In \secref{secObs} we present the VLT/X-shooter observations of SDSS J0831+0354. In \secref{secFit} we identify the absorption troughs of the outflow and measure the ionic column densities which are used to determine the number density (\secref{secFitSIVTroughs}) and the photoionization solution (\secref{secPhoto}). In \secref{secResults} we determine the outflow distance and energetics and compare them to outflows from other objects. We summarize our method and findings in \secref{secSummary}.

\section{OBSERVATIONS AND DATA REDUCTION}\label{secObs}
SDSS J0831+0354 (J2000: RA=08 31 26.15, DEC=+03 54 08.0, z=2.0761) was observed with VLT/X-shooter ($R \sim$ 6000--9000, see Paper I for instrument specifications) as part of our program 92.B-0267 (PI: Benn) in January 2014 with a total integration time of 10640s.

We reduced the SDSS J0831+0354 spectra in a similar fashion to those of SDSS J1106+1939 (detailed in Paper I): we rectified and wavelength calibrated the two-dimensional spectra using the ESO Reflex workflow \citep{ballester11}, then extracted one-dimensional spectra using an optimal extraction algorithm and finally flux calibrated the resulting data with the spectroscopic observations of a standard star observed the same day as the quasar. The one-dimensional spectra were then coadded after manually performing cosmic-ray rejection on each spectra. We present the reduced UVB+VIS spectrum of SDSS J0831+0354 in \figref{figSpecPlot}.

\section{SPECTRAL FITTING}\label{secFit}
\begin{figure*}\begin{minipage}{\textwidth}\includegraphics[width=\textwidth,angle=0]{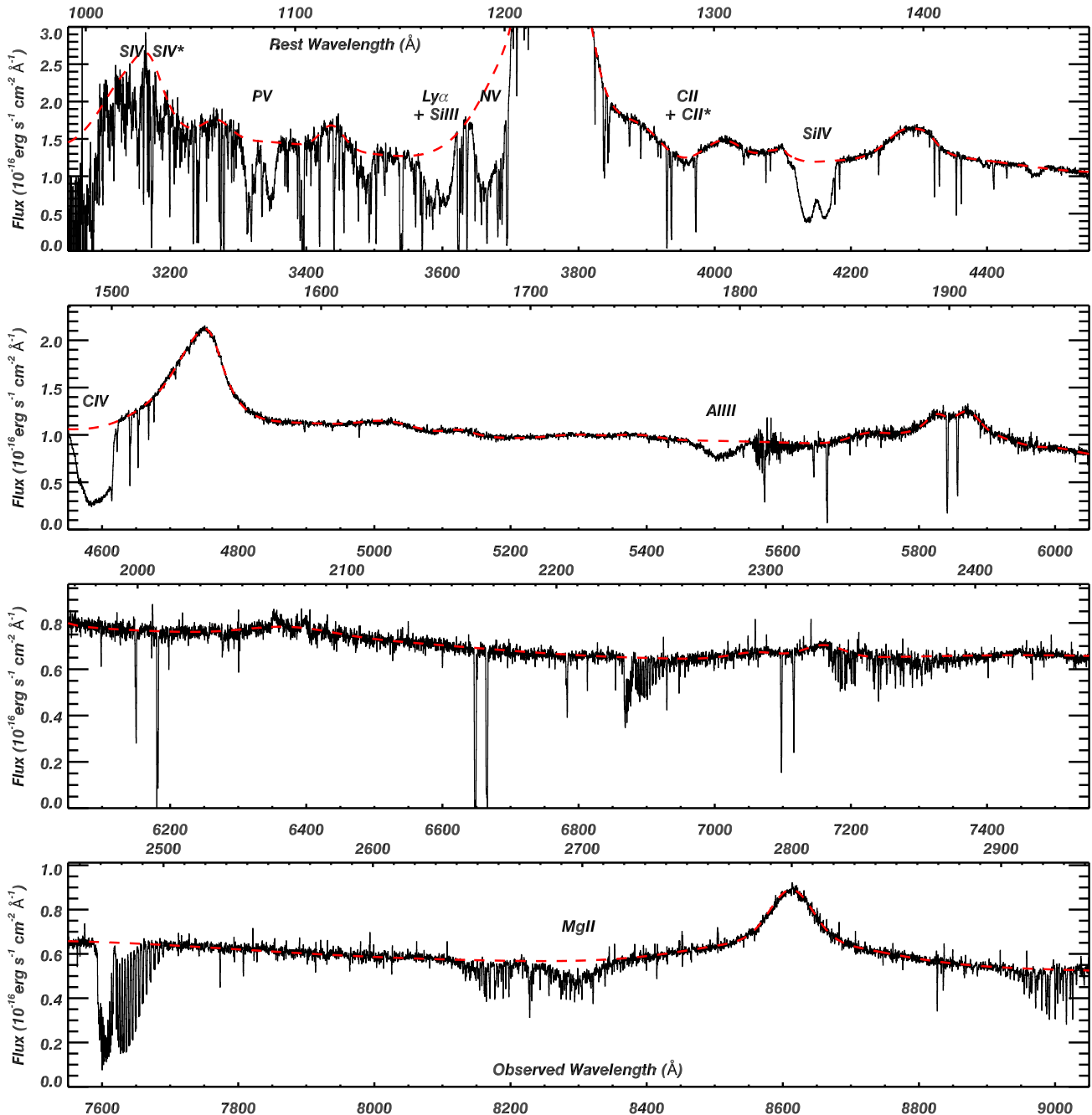}
\caption{VLT/X-shooter spectrum of the quasar SDSS J0831+0354 (z=2.0761). We label the ionic absorption troughs associated with the outflow, and represent the unabsorbed emission model with the red dashed line (see \secref{secFitEmit}). Narrow absorption from intervening systems appear throughout the spectrum, and terrestrial absorption from molecular O$_2$ in our atmosphere is seen near 6850\AA\ and 7600\AA\ observed frame, but none of these features affect the analysis presented here.}
\label{figSpecPlot}\end{minipage}\vspace{-0.3cm}\end{figure*}

Absorption troughs associated with \hi, \civ, \nv, \mgii, \aliii, \siiii, \siiv, \pv\ and \siv/\siv* ionic species are seen in the spectrum (see \figref{figSpecPlot}). The high velocity ($v$=$-$10800 \kms) and width ($3500$ \kms) of the \civ\ trough satisfies the definition of a BAL outflow \citep[i.e. $v_\mathrm{max}>5000$ \kms\ and width $>2000$ \kms, see][]{Weymann81}. \revise{The balnicity index of the \civ\ trough is 1100 \kms\ \citep[see][]{Weymann91}}.
%The high velocity ($v$=$-$10800 \kms), width ($3500$ \kms) and balnicity index (1100 \kms) of the \civ\ trough satisfies the definition of a BAL outflow i.e. $v_\mathrm{max}>5000$ \kms\ and width $>2000$ \kms\ \citep[see][]{Weymann81} and balnicity index $>0$ \kms\ \citep[see][]{Weymann91}.
%\citep[i.e. $v_\mathrm{max}>5000$ \kms, width $>2000$ \kms\ and has a balnicity index of 1100\kms, see][]{Weymann81,Weymann91}
%The high velocity ($v$=$-$10800 \kms) and width ($3500$ \kms) of the \civ\ trough satisfies the definition of a BAL outflow \citep[i.e. $v_\mathrm{max}>5000$ \kms\ and width $>2000$ \kms, see][]{Weymann91}.

\subsection{Unabsorbed Emission Model}\label{secFitEmit}
We correct the spectrum for galactic extinction \citep[$E(B-V)=0.025$;][]{Schlegel98} using the reddening curve of \citet{Cardelli89}. We then fit the continuum with a cubic spline resembling a power law of the form $F(\lambda)=F_{1100} (\lambda/1100)^\alpha$, where $F_{1100}=1.58 \times 10^{-16} \mathrm{erg~s}^{-1} \mathrm{cm}^{-2}$\AA$^{-1}$ is the observed flux at 1100\AA\ (rest-frame) and $\alpha\simeq-1.16$. We model the broad emission lines (BEL) with a sum of one to three Gaussians, which does not significantly affect the column density extraction for most ions since the high velocity of the outflow ($v$=-10800 \kms) shifts the BALs far from the wings of their corresponding BELs. However, the \siv\ and \siv* BALs lie within the \ovi\ BEL, which we model by scaling the \civ\ BEL template to match the peak emission of the \siv\ BAL region.

\subsection{The Blended Troughs}\label{secFitDCTroughs}
We model the optical depth of the absorption troughs with three Gaussians
\begin{equation}\label{eqnGaussianBlend}
	\tau(v)=\sum\limits_{i=1}^3 \tau_i\, \mathrm{exp}\left[\frac{-(v-v_i)^2}{2\sigma_i^2}\right]\hspace{0.5cm}\mathrm{{\small FWHM}}=2\sqrt{2\ln 2}\:\sigma
\end{equation}
 with the same centroid and width in all the observed ionic troughs.  Our first Gaussian ($v_1$=$-$12,100 \kms, {\small FWHM}=1460 \kms) is chosen specifically to match the blue wing of the \siiv\ $\lambda$1393.75 line, as well as the blue wing of the \aliii\ $\lambda$1854.72 line.  The second Gaussian ($v_2$=$-$10,800 \kms, {\small FWHM}=940 \kms), which represents the majority of the absorption, targets the \siiv\ and \pv\ doublets.  A third Gaussian ($v_3$=$-$10,000 \kms, {\small FWHM}=590 \kms) is needed to match the red wing of the \aliii\ $\lambda$1862 line, as well as the red wing of the \civ\ and \nv\ blends.  These three Gaussians form the template used to extract the column density (following the same procedure as Paper I) from the observed troughs of all ions except for \pv\ and \siiii, which will be discussed in the next section. For the three-Gaussian template, we use a covering factor ($C_\mathrm{ion}$) that is constant throughout the trough (i.e. for all three Gaussians), and varies only between ions. For a description of the partial covering (PC) model, see \cite{Arav08}. For ions detected by a singlet, we take the apparent optical depth (AOD) measurement as a lower limit.

\begin{figure*}\begin{minipage}{\textwidth}\includegraphics[height=\textwidth,angle=90,clip=false]{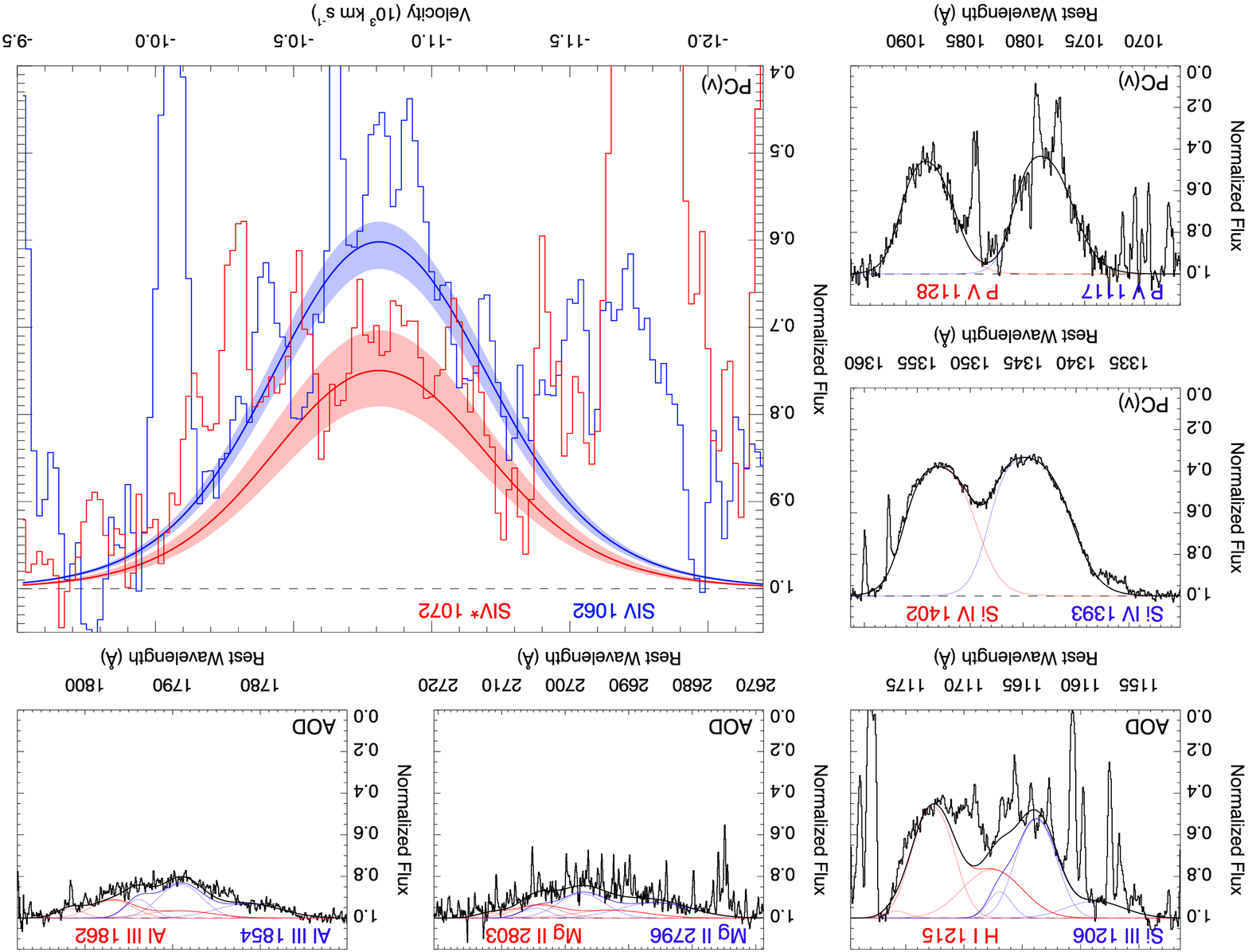}
\vspace{-0.3cm}
\caption{Fits to the absorption troughs observed in the X-shooter spectrum of SDSS J0831+0354.  Each trough is fit by scaling the optical depths of three template Gaussians whose centroid and width are fixed among the ions (see \secref{secFitDCTroughs}). The Gaussian templates for each transition are shown in the same color (red or blue), and the combined absorption from blended lines is represented as a black solid line. \revise{The first row of plots show the apparent optical depth (AOD) fits to \siiii\ (and \hi), \mgii\ and \aliii}. Due to the influence of non-black saturation in the \siiv\ and \pv\ troughs (see text), we show the velocity-dependent partial covering (PC(v)) fit to those ions. \revise{The plot in the lower-right shows the PC(v) fit to the \siv\ and \siv* troughs in velocity space. The fits are spanned by shaded contours representing the upper and lower errors assigned to the fits.}}
%\caption{Fits to the absorption troughs observed in the X-shooter spectrum of SDSS J0831+0354.  Each trough is fit by scaling the optical depths of three template Gaussians whose centroid and width are fixed among the ions (see \secref{secFitDCTroughs}). The Gaussian templates for each transition are shown in the same color (red or blue), and the combined absorption from blended lines is represented as a black solid line. Due to the influence of non-black saturation in the \siiv\ and \pv\ troughs (see text), we show the velocity-dependent partial covering (PC(v)) fit to only those ions and to \siv; all others are apparent optical depth (AOD) fits.}
\label{figVCuts}\end{minipage}\vspace{-0.3cm}\end{figure*}

\subsection{Velocity-dependent covering: \pv\ and \siiii}\label{secFitVCovTroughs}
Although the central Gaussian of our template (see \secref{secFitDCTroughs}) can be used to model the \pv\ and \siiv\ doublet troughs, the ratio between the red and the blue components of each doublet approaches 1:1. This suggests that the troughs are strongly influenced by the velocity-dependent partial covering of the emission source \cite[see][]{Borguet12b,Arav99b}. 
Due to the presence of \Lya\ forrest intervening troughs, we proceed by modeling the normalized flux of the doublets with a smooth function.  For phenomenological flexibility, we built a function from the product of two logistic functions (hereafter logistic fit)
\begin{equation}\label{eqnLogisticDef}
 I(v)=1-\frac{A}{\left(1+e^{w_b(v-v_b)}\right)\left(1+e^{w_r(v-v_r)}\right)}
\end{equation}
where $A$ is the maximum depth of the logistic fit, $v_b$ and $v_r$ are the velocities of the blue and red wing half-maximums respectively ($v_b\le v_r$), and $w_b$ and $w_r$ are the slope/variance of the blue and red wings respectively ($w_b<0<w_r$). This function is particularly suited to fitting asymmetric troughs.

We model each \pv\ doublet component (normalized and in velocity-space) using a logistic fit, and determine the velocity-dependent covering fraction and optical depth solution \cite[using Equation (3) of][]{Dunn10a}. Integrating this optical depth over the trough yields the \pv\ column density for the partial covering model given in \tabref{tabColDen}.

\siiii\ is a singlet, thus we cannot determine a covering model for this ion. Therefore, we use the covering factor of \siiv\ as a proxy to extract the \siiii\ PC column density given in \tabref{tabColDen}, since \siiv\ is the next ionization stage of the same element. The lower error of 0.11 dex for the adopted \siiii\ measurement includes systematic errors in continuum placement and blending of the intervening Ly$\alpha$ forest.  We note that, by coincidence, the difference between the PC measurement and the apparent optical depth (AOD) measurement is also 0.11 dex.

%\siiii\ is a singlet, thus we cannot determine a covering model for this ion.  Like \pv, \siiv\ exhibits a high degree of saturation, and we perform a similar fit to the \siiv\ doublet and extract the velocity-dependent partial covering model.  We then assume that \siiii\ has the same covering model, and use the \siiv\ template to extract the \siiii\ PC column density given in \tabref{tabColDen}.  The lower error of 0.11 dex for the adopted \siiii\ measurement includes systematic errors in continuum placement and blending of the intervening Ly$\alpha$ forest.  We note that, by coincidence, the difference between the PC measurement and the (lower limit) AOD is also 0.11 dex.

\subsection{The density-sensitive troughs: \siv\ and \siv*}\label{secFitSIVTroughs}

The column densities from \siv\ and its excited state \siv* play an essential role in our analysis: the ratio of level populations between \siv* and \siv\ is dictated by collisional excitation, which depends on \ne\ and is relatively insensitive to temperature (see Figure 8 of Paper I).

%\revise{We use two main methods to determine the column density of \siv\ and \siv*: PC(v) and AOD (see \tabref{tabColDen}).}
As in the case of \siiii, \siv\ $\lambda$1062.66 and \siv* $\lambda$1072.97 are both singlets (the lines arise from different energy levels with a priori undetermined populations), thus a velocity dependent covering solution cannot be determined \revise{from these two lines alone}.
\revise{Therefore, we follow the same method as with \siiii, using the velocity-dependent covering template of \siiv\ to model the \siv/\siv* troughs with the PC(v) method.}

\revise{
We begin this method by modeling the \siv/\siv* troughs using a logistic curve that closely resembles the central Gaussian of the three Gaussian template.
This resemblence ensures that the \siv\ absorption model is kinematically related to the other ionic troughs in the outflow.
The logistic curve was fit to the \siv\ troughs by scaling the amplitude until it fit the broad absorption features in the spectrum while excluding the narrow Ly$\alpha$ troughs.
To assign the upper and lower errors to this fit we again scale the depth of the logistic template until the resulting template clearly over- or under-predicts significant portions of the broad absorption trough.
The quality of this fit and of the stated errors can be judged in the lower right panel of \figref{figVCuts}.
}

\revise{
The optical depth of the fit (and its errors) was translated to \siv/\siv* column density measurements using the same procedure as in \siiii\ (i.e. adopting the velocity-dependent covering solution for \siiv).
These column densities are reported in the fourth column of \tabref{tabColDen}.
%This is the same procedure that was used to fit the \pv\ troughs that are affected by Ly$\alpha$ forest absorption features.
A similar value (differing by only 0.05 dex) was obtained with the \pv\ velocity-dependent covering template, demonstrating the robustness of this method.
}

\revise{
To obtain the AOD column density, we follow the same procedure as the PC(v) modeling, instead using the central Gaussian of the three Gaussian template described in \secref{secFitDCTroughs}.
For completion, we also apply the \emph{constant} partial covering template of \siiv\ to the \siv/\siv* absorption (using the central Gaussian as the optical depth template), and report the PC column density in the third column of \tabref{tabColDen}
}

%\revise{To obtain the AOD column density, we follow the same procedure as the PC(v) modeling, instead using the central Gaussian of the three Gaussian template described in \secref{secFitDCTroughs}.}

%\revise{Finally, using the same method as for \siiii, we also apply the constant partial covering template of \siiv\ to the \siv/\siv* absorption (using the central Gaussian as the optical depth template), and obtain the values reported in the third column of \tabref{tabColDen}}

\begin{figure}\includegraphics[width=\columnwidth,angle=0]{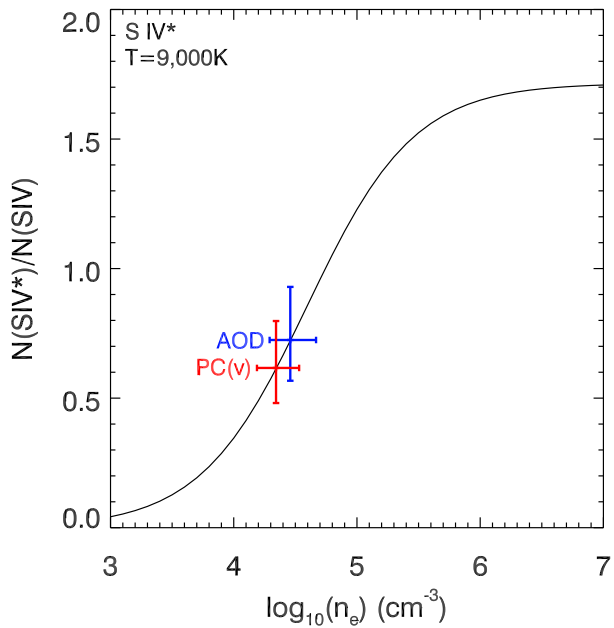}
\caption{Density diagnostic using the \siv*$/$\siv\ ratio for the SDSS J0831+0354 outflow, presented in the same manner as Paper I: we plot the \siv*$/$\siv\ theoretical ratio versus electron number density and overlay our measured value (for two absorber models, AOD and \revise{velocity-dependent PC(v)}) to determine the number density and its errors.}
\label{figExFrac}\vspace{-0.4cm}\end{figure}

We deduce the \ne\ of the outflow by comparing the \siv*$/$\siv\ column density ratio to predictions (see \figref{figExFrac}) made with the Chianti 7.1.3 atomic database \citep{Landi13}. Using the column densities for \siv\ and \siv* reported in \tabref{tabColDen} and the electron temperature found for our best-fit Cloudy models from \secref{secPhoto} (9000~K, a weighted average for \siv\ across the slab), we find $\log(n_\mathrm{e}) = 4.42^{+0.26}_{-0.22}$ cm$^{-3}$ by averaging the two absorption models (reported in \tabref{tabne}) and adding their errors in quadrature.

\begin{table}
\caption{SDSS J0831+0354 log column densities (cm$^{-2}$) for the three absorption models AOD, PC and PC(v) (velocity-dependent covering). Upper limits are shown in red and lower limits are in blue.}
\label{tabColDen}\begin{tabular}{r@{$\,$}lr@{}c@{}lr@{}c@{}lr@{}c@{}lr@{}c@{}l}\hline
%\multicolumn{14}{l}{{\bf Table 1:} J0831 Column densities}\\\hline
\multicolumn{2}{c}{Ion}	&\multicolumn{3}{c}{AOD}	&\multicolumn{3}{c}{PC}		&\multicolumn{3}{c}{PC(v)}		&\multicolumn{3}{c}{Adopted$^a$}	\\\hline
H&{\sc i}		&\LL{14.96}			&\LL{15.17}			&&-----&			&\LL{15.17}			\\
He&{\sc i}$^*$		&\UL{15.14}			&&-----&			&&-----&			&\UL{15.14}			\\
C&{\sc ii}		&\UL{14.46}			&&-----&			&&-----&			&\UL{14.46}			\\
C&{\sc iv}		&\LL{15.50}			&\LL{15.65}			&&-----&			&\LL{15.65}			\\
N&{\sc v}		&\LL{15.60}			&\LL{15.79}			&&-----&			&\LL{15.79}			\\
Mg&{\sc ii}		&\LL{13.78}			&\Meas{13.87}{0.13}{0.13}	&&-----&			&\Meas{13.87}{0.13}{0.13}	\\
Al&{\sc ii}		&\UL{13.07}			&&-----&			&&-----&			&\UL{13.07}			\\
Al&{\sc iii}		&\LL{14.10}			&\Meas{14.27}{0.13}{0.13}	&&-----&			&\Meas{14.27}{0.13}{0.13}	\\
Si&{\sc iii}		&\LL{14.27}			&\LL{14.35}			&\LL{14.38}			&\color{blu}$>$&\color{blu}14.38&\color{blu}$^b_{-0.11}$\color{blk}	\\
Si&{\sc iv}		&\LL{15.17}			&\LL{15.44}			&\LL{15.40}			&\LL{15.40}			\\
P&{\sc v}		&\LL{15.18}			&\Meas{15.33}{0.05}{0.05}	&\Meas{15.47}{0.15}{0.10}	&\Meas{15.47}{0.15}{0.10}	\\
S&{\sc iv}		&\LL{15.53}			&\Meas{\revise{15.48}}{0.07}{0.10}	&\Meas{15.60}{0.04}{0.05}	&\Meas{\revise{15.60}}{0.04}{0.05}	\\
S&{\sc iv}$^*$		&\LL{15.35}			&\Meas{\revise{15.40}}{0.07}{0.12}	&\Meas{15.39}{0.10}{0.10}	&\Meas{\revise{15.39}}{0.10}{0.10}	\\\hline
\multicolumn{14}{l}{\parbox[t]{\columnwidth}{$^a$Adopted value for photoionization modelling (see text).}}\\
\multicolumn{14}{l}{\parbox[t]{\columnwidth}{$^b$We use the difference between the \siiii\ AOD and PC measurements as a lower error on the \siiii\ lower limit.}}\\
\end{tabular}\vspace{-0.05cm}\end{table}

\begin{table}
\caption{Number density measurements. The values correspond to the horizontal error-bars in \figref{figExFrac}.}
\label{tabne}\begin{tabular}{rrrr}\hline
			&AOD			&\revise{PC(v)}			&Adopted\\\hline
$\log(n_\mathrm{e})$	&$4.47^{+0.21}_{-0.18}$	&$4.36^{+0.19}_{-0.16}$	&$4.42^{+0.26}_{-0.22}$\\\hline
\end{tabular}\vspace{-0.05cm}\end{table}

\section{PHOTOIONIZATION ANALYSIS}\label{secPhoto}
We use photoionization models in order to determine the ionization equilibrium of the outflow, its total hydrogen column density (\Nh), and to constrain its metallicity. The ionization parameter
\begin{equation}\label{eqnUDef}
 U_\mathrm{H}\equiv\frac{Q_\mathrm{H}}{4 \pi R^2 c n_\mathrm{H}},
\end{equation}
(where \qh\ is the source emission rate of hydrogen ionizing photons, $R$ is the distance to the absorber from the source, $c$ is the speed of light, and \nh\ is the hydrogen number density) and \Nh\ of the outflow are determined by self-consistently solving the ionization and thermal balance equations with version c08.01 of the spectral synthesis code Cloudy, last described in \citet{Ferland13}. We assume a plane-parallel geometry for a gas of constant \nh\ and initially choose solar abundances and the UV-Soft SED which is a good representation of radio-quiet quasars \citep[described in Section~4.2 of][other SEDs and metallicities will be explored in \secref{secPhotoSEDZ} here]{Dunn10a}. For the chosen SED and metallicity, we generate a grid of models by varying \Nh\ and \uh. Ionic column densities ($N_\mathrm{ion}$) predicted by the models are tabulated and compared with the measured values in order to determine the models that best reproduce the measured $N_\mathrm{ion}$. For a more elaborate description of this method, see \cite{Borguet13,Arav13}.

\subsection{Photoionization Solution}\label{secPhotoSoln}
\begin{figure}\includegraphics[width=\columnwidth,angle=0]{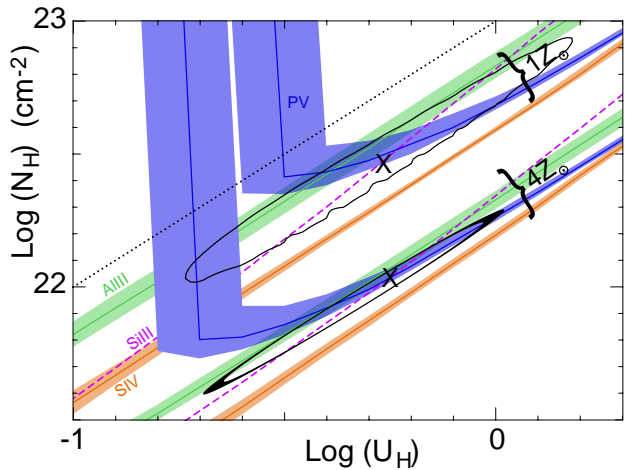}
\caption{Phase plot showing the photoionization solution using the UV-Soft SED for gas with one and four times solar metallicity. Each colored contour represents (for one of the two metallicities) the locus of models $\left(U_{H},N_{H}\right)$ which predict a column density consistent with the observed column density for that ion.  The bands which span the contours are the 1-$\sigma$ uncertainties in the measured observations. The dashed line indicates the \siiii\ lower limit.  For each metallicity, the black ``X'' is the best ionization solution and is surrounded by the $\chi^2$ contour (see text).}
\label{figPhasePlot}\vspace{-0.0cm}\end{figure}

In \figref{figPhasePlot}, contours where model predictions match the measured $N_\mathrm{ion}$ are plotted in the \Nh$-$\uh\ plane.  Note that \figref{figPhasePlot} presents models for two different metallicity values (1\Zsun\ and 4\Zsun, see discussion in \secref{secPhotoSEDZ}).  The ionic column densities used in the photoionization modelling are listed in the last column of \tabref{tabColDen}, consisting of the PC measurements where available and with PC(v) prioritized over PC.  In \figref{figPhasePlot}, we show only the ions which dominate the solution; the lower limits (\hi, \civ, \nv\ and \siiv) and upper limits (\hei*, \cii\ and \alii) are trivially satisfied by the contraints of \aliii\ and \siv.  The \mgii\ contour lies between the \aliii\ and \siv\ contours, and is thus also satisfied. 

For $Z=1\Zsun$, the \pv\ contour requires a higher \Nh\ than the \siv\ contour in regions also satisfied by the \siiii\ lower limit (above the dashed line in \figref{figPhasePlot}). This over-prediction of the \siv\ column density will be discussed in \secref{secResultsSIVDiscrepancy}. The solution is also influenced by the \aliii\ contour which lies parallel to \siv. This results in a poor fit ($\chi^2_\mathrm{red}$=10.6), since no models can simultaneously predict the observed \siv\ and \aliii\ column denisties.

%For $Z=1\Zsun$, the \siiii\ lower limit drives the solution above the dashed line in \figref{figPhasePlot}, which favors models with a higher column density or lower ionization parameter.  The \pv\ contour complements the \siiii\ lower limit, requiring a higher column density with higher ionization parameter.  These two contraints, along with the \aliii\ measurement, drive the solution above the \siv\ contour, over-predicting the observed \siv\ column density by a factor of three.  This discrepancy also arises from the parallel contours of \aliii\ and \siv, which are responsible for the elongation of the $1\sigma$ confidence interval (black contours in \figref{figPhasePlot}, see \citealt{Arav13} for definition).

\begin{figure}
\includegraphics[height=\columnwidth,angle=90]{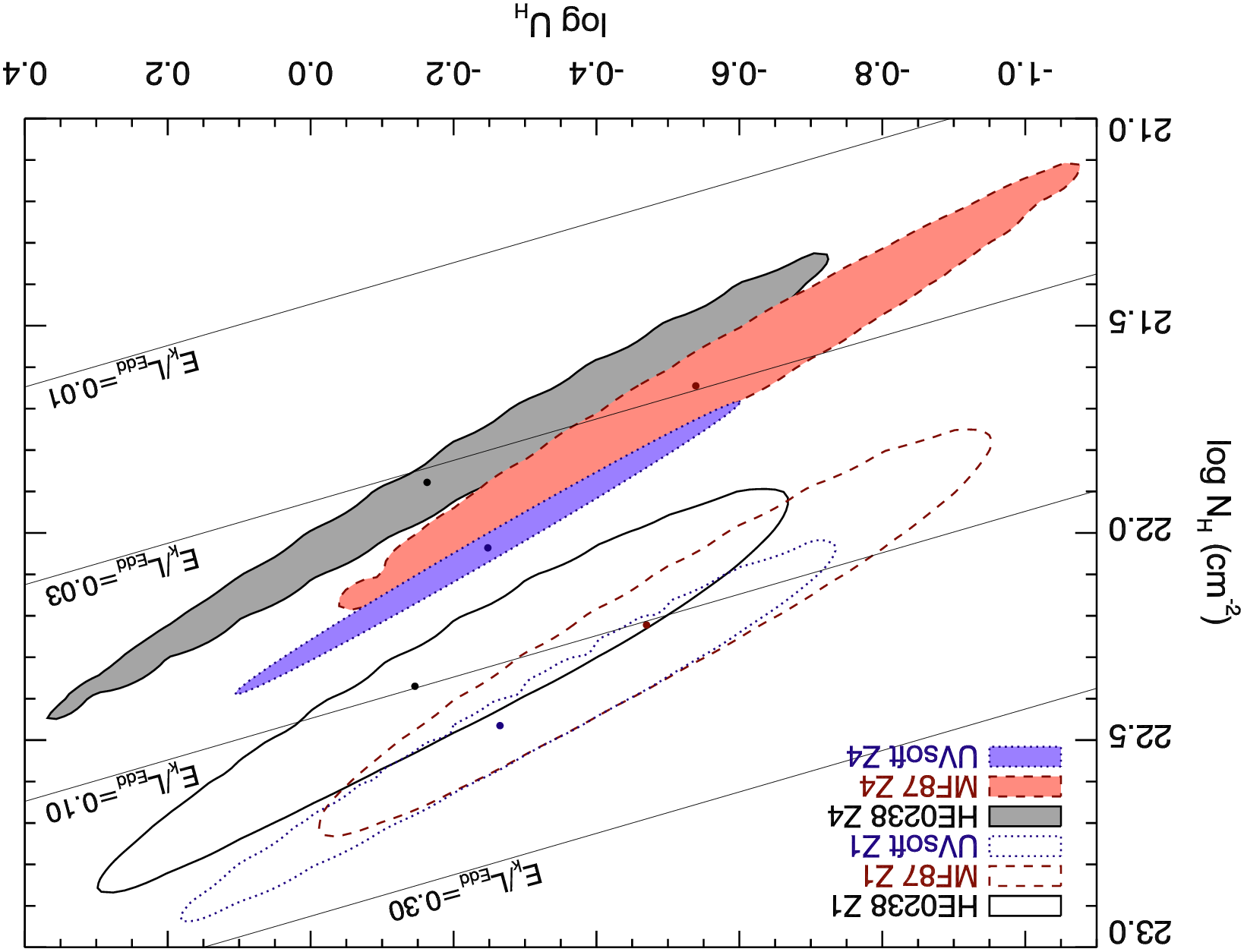}\\
\includegraphics[height=\columnwidth,angle=90]{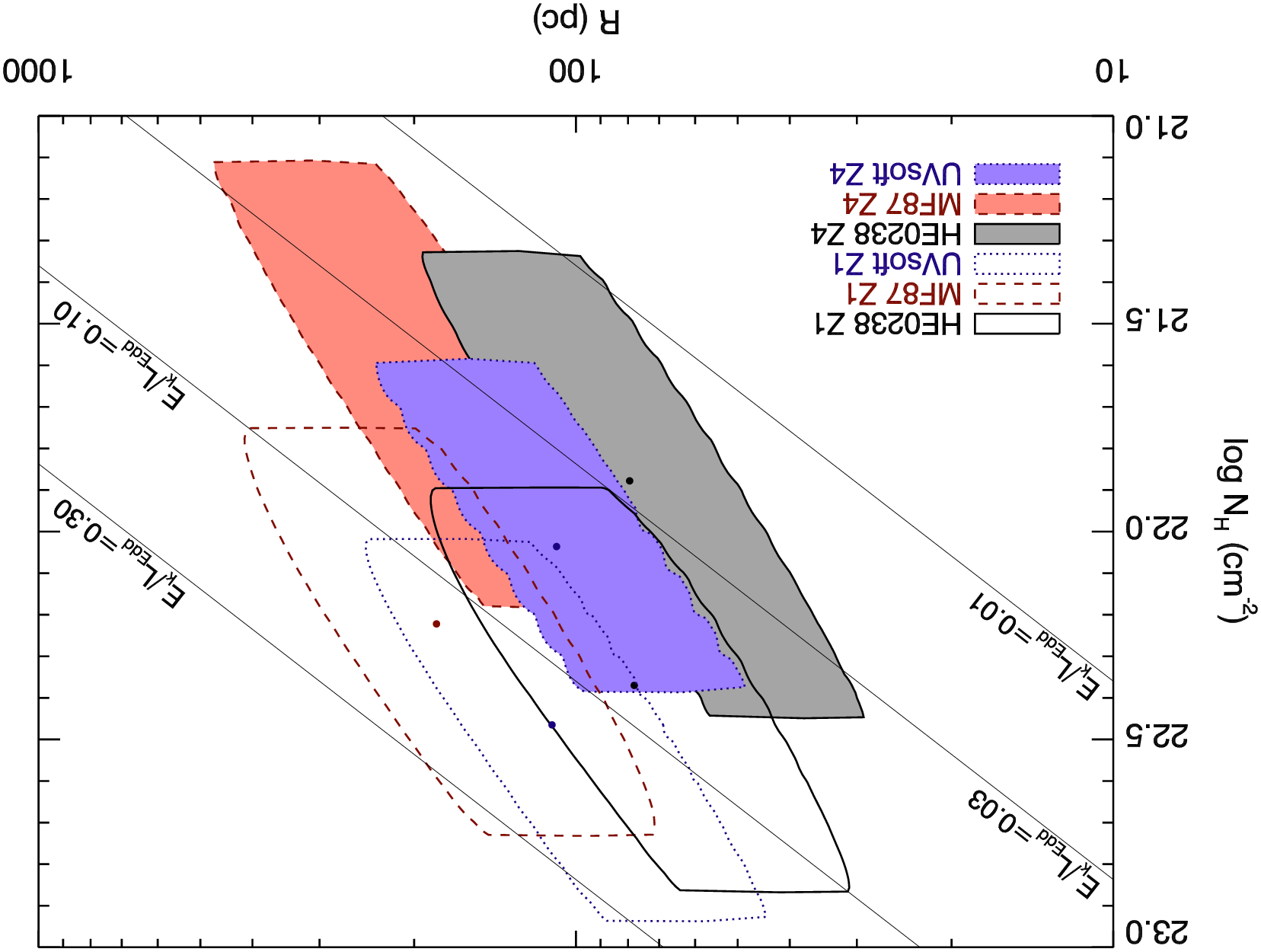}
\caption{{\bf(top)} Phase plot showing the photoionization solution for three SEDs (HE0238, MF87 and UVsoft, see descriptions in \citealt{Arav13}) and two metallicities (one solar (Z1) and four solar (Z4)) for a total of six models. Contours of equal \edot$/L_\mathrm{Edd}$ (assuming fixed \nh) are shown as parallel, thin solid lines assuming the UV-Soft SED (see text). {\bf(bottom)} Same as above, but with the horizontal axis converted to distance, which incorporates the errors on \nh\ and \uh\ into the horizontal errors on $R$.}
\label{figMultiModel}\vspace{-0.5cm}\end{figure}

\subsection{Dependence on SED and metallicity}\label{secPhotoSEDZ}
%The discrepancy between the \pv\ and \siv\ contours can be alleviated by considering a higher metallicity, as outflows are known to have supersolar metallicities \citep[e.g.][]{Gabel06,Arav07}.  \figref{figPhasePlot} also shows the photoionization solution for a metallicity of $Z=4\Zsun$, scaled according to the Cloudy starburst schema \cite[following grid M5a of][]{Hamann93}. This model scales the abundances of the elements we observe by the same factor, and results in only a slightly better fit (for $Z=4\Zsun$, $\chi^2_\mathrm{red}$=7.2), although the required \Nh\ of the cloud decreases by 0.4 dex. We will discuss alternative elemental abundances in \secref{secResultsSIVDiscrepancy}; here we consider the same metallicities used for other outflows studied by our group (see next paragraph).

%The SED is poorly constrained throughout the ionizing portion of the quasar spectrum \citep{Dunn10a}; thus we consider different SEDs in our photoionization analysis. The chemical composition of the outflowing gas is likewise a source of systematic uncertainty, as outflows are known to have differing values of supersolar metallicities \citep[e.g.][]{Gabel06,Arav07}.

We consider the sensitivity of the photoionization solution to our choice of SED and metallicity by following the approach of \cite{Arav13}, thereby allowing for comparison with previous outflows under the same set of assumptions.  We find the photoionization solution in six different cases (three SEDs from \citealt{Arav13} each with $1\Zsun$ and $4\Zsun$), plot the $\chi^2$ contours in the top panel of \figref{figMultiModel} and report the solutions in \tabref{tabEnerCompare}. The change in metallicity from $1\Zsun$ to $4\Zsun$, scaled according to the Cloudy starburst schema \cite[following grid M5a of][]{Hamann93}, has the same effect with each SED, decreasing the log(\Nh) of the solution by 0.4 dex. This is illustrated in \figref{figPhasePlot}, which shows the photoionization solution for metallicities of both $Z=1\Zsun$ and $Z=4\Zsun$. We will discuss alternative elemental abundances in \secref{secResultsSIVDiscrepancy}; here we consider the same metallicities used for other outflows studied by our group.

\begin{table}
\caption{Properties of the chosen SEDs.}
\label{tabQHLBol}\begin{tabular}{lrrr}\hline
SED				&HE0238		&MF87		&UV-Soft\\\hline
log(\lbol) (erg s$^{-1}$)	&46.79		&47.03		&46.87\\
log(\qh) (s$^{-1}$)		&56.53		&56.95		&56.72\\\hline
\end{tabular}\vspace{-0.3cm}\end{table}

The different SEDs \revise{we use} spread the solution over 0.4 dex in log(\uh), which is comparable to the errors on each solution. In \tabref{tabQHLBol} we report \qh\ as well as the bolometric luminosity \lbol\ for the three SEDs by fitting them to the measured flux (corrected for Galactic reddening) at 1100~\AA\ (in the rest-frame) and integrating over the whole energy range.

One important motivation to find both the ionization parameter \uh\ and the number density \nh\ is to determine the distance of the outflow $R$ from \eqnref{eqnUDef}. We therefore convert the ionization parameter log(\uh) to distance $R$ using \eqnref{eqnUDef} and present the same six photoionization solutions in the bottom panel of \figref{figMultiModel}. The uncertainties on \uh\ and \nh\ are incorporated into our value for $R$, which forms the horizontal spread of the contours in \figref{figMultiModel}. Since \Nh\ is not involved in the distance determination, the vertical spread of the contours in \figref{figMultiModel} remains unchanged. In both panels of \figref{figMultiModel}, we also show the contours of equal \edot$/L_\mathrm{Edd}$ (see \secref{secResultsEner}) assuming the UV-Soft SED, noting that the same contours for the different SEDs are shifted in log(\Nh) by no more than 0.1 dex.

\section{RESULTS AND DISCUSSION}\label{secResults}

\subsection{Energetics}\label{secResultsEner}
Assuming the outflow is in the form of a thin partial shell, its mass flow rate (\mdot) and kinetic luminosity (\edot) are given by \citep[see][for discussion]{Borguet12a}
\begin{eqnarray} \dot{M} &=& 4 \pi R \Omega \mu m_p N_{H} v \label{mdoteq} \\ \dot{E}_\mathrm{k} &=& 2 \pi R \Omega \mu m_p N_{H} v^3 \label{ekeq} \end{eqnarray}
where $R$ is the distance from the outflow to the central source, $\Omega$ is the global covering fraction of the outflow, $\mu$ = 1.4 is the mean atomic mass per proton, $m_p$ is the mass of the proton, $N_H$ is the total hydrogen column density of the absorber, and $v$ is the radial velocity of the outflow. Using the parameters reported in the preceeding sections, we calculate the energetics for the six models we have considered for the \siv\ outflow of SDSS J0831+0354 and report the relevant values in the first six rows of \tabref{tabEnerCompare} (the other four rows are comparison outflows that will be discussed in \secref{secResultsComparison}). We adopt the UV-Soft SED with $Z=4\Zsun$ as our representitive model for this outflow, which is the same model chosen in Paper I, and yields a conservative measurement of \edot. As in Paper I, we use $\Omega=0.08$, which is appropriate for \siv\ BAL outflows.

%The feedback criteria derived in \cite[sections 2 and 3 of][]{Scannapieco04} expresses \edot\ as a fraction of the Eddington luminosity ($L_\mathrm{Edd}$) which is not necessarily equal to the Bolometric luminosity ($L_\mathrm{Bol}$) of the quasar. Using $L_\mathrm{Edd}$ instead of $L_\mathrm{Bol}$ for this purpose better reflects the growth of the SMBH and its long-term connection with the host galaxy, since the black hole shines at $L_\mathrm{Edd}$ during the periods of its most rapid growth.

As noted in the Introduction, an \edot\  value of at least 0.5\% \citep{Hopkins10} or 5\% \citep{Scannapieco04} of the Eddington luminosity (\ledd) is deemed sufficient to produce significant AGN feedback effects. 
 We determine the Eddington luminosity $L_\mathrm{Edd}=10^{\revise{47.0}}$~erg~s$^{-1}$ from the mass of the SMBH (e.g. using Equation~(6.21) of \cite{KrolikBook}). The mass of the SMBH is determined using the virial mass estimator from Equation~(\revise{3}) of \revise{\cite[][given the caveat that the scaling relationship extrapolated to high-z/luminosity quasars is not yet firmly established]{Park13}. This estimate} requires the continuum luminosity at rest-frame 1350\AA\ and the FWHM of the \civ\ BEL, both of which are directly measured from our spectrum. We use this method to determine $L_\mathrm{Edd}$ for the quasars SDSS J1106+1939 and SDSS J0838+2955. The \civ\ BEL in SDSS J0318-0600 is not well-defined, so we perform a similar analysis using the \mgii\ BEL \revise{and Equation~(1) of \cite{Vestergaard09}}. The only spectral coverage of HE 0238-1904 is in the extreme-UV, which does not cover any diagnostic lines sufficient for this method. We therefore use the assertion \citep[from Section 9 of][]{Arav13} that the Eddington luminosity is approximately the bolometric luminosity for our comparison in \tabref{tabEnerCompare2}.

\revise{We note that recent studies \citep{Luo14} have found X-ray weak SEDs for BALQSO that are much softer than the UVsoft SED we use here. To explore the effects of such an SED on our results, we constructed an SED from the object in \cite{Luo14} showing the most extreme X-ray softness (PG 1254+047) by interpolating between the given flux points \citep[see right-centre panel of Figure 3 of][]{Luo14}. At 2 keV the flux is three orders of magnitude lower than in our UVsoft SED. The photoionization solution for this very soft SED leads to a distance of 51 pc and \edot=10$^{44.9} \mathrm{erg~s}^{-1}$ for the outflow using Z=4\Zsun. This distance and energy is lower by a factor of two and six respectively, compared to the values obtained with the UVsoft SED (110 pc, \edot=10$^{45.7} \mathrm{erg~s}^{-1}$) for Z=4\Zsun.}

\revise{We caution that the cause of the observed X-ray weakness can be attributed to optically thick absorption from the outflow itself, rather than intrinsically weak X-ray emission or obscuration between the central source and the outflow. If this is the case, then the SED incident on the outflow would likely resemble one of the other SEDs presented here. It is the incident radiation that will determine the photoionization solution.}

%The cause of the observed X-ray weakness can be attributed to optically thick absorption from the outflow rather than intrinsically weak X-ray emission incident on the outflow. If this is the case, then the SED before attenuation (incident on the outflow) would likely resemble one of the SEDs we have selected. Considering that the \Nh\ we determined for the outflow in SDSS J0831+0354 is high enough to form a \heii\ ionization front, heavy obscuration of the quasar at energies greater than 4 Ryd is certainly possible. However, our photoionization analysis requires the SED incident on the outflow, thus an X-ray weak SED resulting from attenuation would not be applicable to our analysis.

\begin{table*}
\begin{minipage}{\textwidth}
\caption{Physical properties of energetic quasar outflows (object names abbreviated from \tabref{tabEnerCompare2}).}
\label{tabEnerCompare}
\begin{tabular*}{\textwidth}{@{\extracolsep{\fill}}lrrrrrrrrr}\hline
Object, SED	&\hspace{-0.3cm}log(\lbol)	&v		&$\log\left(U_H\right)$	&$\log\left(N_H\right)$		&log($n_\mathrm{e}$)	&$R$			&$\dot{M}$						&$\log(\dot{E}_k)$				&$\dot{E}_k$/$L_{\mathrm{Edd}}$\\
		&\hspace{-0.3cm}(erg s$^{-1}$)	&(km s$^{-1}$)	&			&($\mathrm{cm}^{-2}$)		&(cm$^{-3}$)		&(pc)			&$\left(\mathrm{M}_\odot \mathrm{yr}^{-1}\right)$	&(erg s$^{-1}$)					&(\%)\\\hline
J0831 HE0238 Z1 &46.8 &-10800 &$-0.15^{+0.4}_{-0.5}$ &$22.4^{+0.5}_{-0.5}$ &$4.42^{+0.26}_{-0.22}$ &$78^{+27}_{-18}$ &$230^{+330}_{-130}$ &$45.9^{+0.4}_{-0.3}$ &\revise{8$^{+11}_{-4.4}$}\\
J0831 MF87 Z1 &47.0 &-10800 &$-0.47^{+0.5}_{-0.5}$ &$22.2^{+0.5}_{-0.5}$ &$4.42^{+0.26}_{-0.22}$ &$180^{+53}_{-44}$ &$390^{+570}_{-220}$ &$46.1^{+0.4}_{-0.4}$ &\revise{13$^{+19}_{-7.5}$}\\
J0831 UVsoft Z1 &46.9 &-10800 &$-0.26^{+0.4}_{-0.5}$ &$22.5^{+0.5}_{-0.4}$ &$4.42^{+0.26}_{-0.22}$ &$110^{+30}_{-25}$ &$410^{+530}_{-220}$ &$46.2^{+0.4}_{-0.3}$ &\revise{14$^{+18}_{-7.7}$}\\
J0831 HE0238 Z4 &46.8 &-10800 &$-0.16^{+0.5}_{-0.6}$ &$21.9^{+0.6}_{-0.6}$ &$4.42^{+0.26}_{-0.22}$ &$79^{+33}_{-24}$ &$76^{+120}_{-46}$ &$45.4^{+0.4}_{-0.4}$ &\revise{2.6$^{+4.2}_{-1.6}$}\\
J0831 MF87 Z4 &47.0 &-10800 &$-0.54^{+0.5}_{-0.5}$ &$21.6^{+0.5}_{-0.5}$ &$4.42^{+0.26}_{-0.22}$ &$200^{+74}_{-54}$ &$110^{+170}_{-66}$ &$45.6^{+0.4}_{-0.4}$ &\revise{3.8$^{+5.8}_{-2.3}$}\\
J0831 UVsoft Z4 &46.9 &-10800 &$-0.25^{+0.3}_{-0.4}$ &$22.0^{+0.4}_{-0.5}$ &$4.42^{+0.26}_{-0.22}$ &$110^{+27}_{-15}$ &$150^{+140}_{-84}$ &$45.7^{+0.3}_{-0.4}$ &\revise{5.2$^{+4.8}_{-2.9}$}\\
J1106 UVsoft Z4 &47.2 		&-8250	 	&$-0.5^{+0.3}_{-0.2}$ 	&$22.1^{+0.3}_{-0.1}$ 		&$4.1^{+0.02}_{-0.37}$	&$320^{+200}_{-100}$    &$390^{+300}_{-10}$  					&$46.0^{+0.3}_{-0.1}$ 				&\revise{12$^{+11}_{-0.3}$}\\
HE 0238-1904 Z4$^a$&47.2		&-5000		&$0.5^{+0.1}_{-0.1}$	&$20.0^{+0.1}_{-0.1}$		&$3.83^{+0.10}_{-0.10}$	&$3400^{+2000}_{-490}$	&$140^{+80}_{-40}$					&$45.0^{+0.2}_{-0.2}$				&$0.7^{+0.5}_{-0.2}$\\
SDSS J0838+2955 &47.5 		&-5000	 	&$-2.0^{+0.2}_{-0.2}$	&$20.8^{+0.3}_{-0.3}$ 		&3.8 			&$3300^{+1500}_{-1000}$	&$300^{+210}_{-120}$					&$45.4^{+0.2}_{-0.2}$ 				&\revise{2.3$^{+1.5}_{-0.8}$}\\
SDSS J0318-0600	&47.7		&-4200		&-3.1			&19.9				&3.3			&6000			&120							&44.8						&0.13\\\hline
\multicolumn{10}{l}{$^a$For the high-ionization phase of trough B.}\\
\end{tabular*}
\end{minipage}\vspace{-0.0cm}
\end{table*}

\subsection{The \siv\ discrepancy}\label{secResultsSIVDiscrepancy}
Our photoionization solution (for the UV-soft SED) over-predicts the \siv\ total column density by a factor of eight, which is not alleviated appreciably by changing the SED or metallicity. We offer two resolutions to this discrepancy: a two-phase solution, or individual abundance scaling of sulphur.

%Our photoionization solution (for the UV-soft SED) over-predicts the \siv\ total column density by a factor of eight, which is not alleviated by changing the SED or metallicity. This suggests that the \siv\ line could be saturated, rendering its column density unusable. \siv* cannot be saturated since its trough is shallower than \siv\ (see discussion in Section 6.3 of Paper I). Since the number density determination hinges soley on the \siv*$/$\siv\ column density ratio (the \cii*$/$\cii\ ratio saturates at 2 for log \ne$>$3), this discrepancy casts doubt on one of the key parameters in our analysis. We offer two resolutions to this discrepancy: a two-phase solution, or individual abundance scaling of sulphur.

1) \cite{Arav13} showed that two ionization phases exist in some outflows (in that case smaller, denser cloudlets embedded in high-ionization gas with much larger \Nh), and can account for the appearence of low-ionization species that are inconsistent with the ionization solution obtained from higher-ionization species. \siv\ and \pv\ are both high-ionization species, thus their column densities can be produced in a separate phase than the low-ionization species (\mgii\ and \aliii). The \siv\ and \pv\ contours in \figref{figPhasePlot} converge at high ionization parameters (log \uh $>$0.3). A high-ionization phase at this ionization parameter would satisfy both the \siv\ and \pv\ constraints, but under-predict the \mgii\ and \aliii\ column densities, which would need to be satisfied by a low-ionization solution. A two-phase solution could resolve the \siv\ discrepancy, but such a solution is not well-constrained with the measurements obtained from our spectra. However, any high-ionization solutions that adhere to the \pv\ constraint would move towards the upper-right corner of \figref{figPhasePlot}. Since $\dot{E}_\mathrm{k}\propto N_\mathrm{H}/\sqrt{U_\mathrm{H}}$, this increase in both \uh\ and \Nh\ by 0.6 dex would result in an increase in \edot\ of 0.3 dex.

2) The discrepancy between \siv\ and the other ions may be alleviated by considering a higher metallicity, as outflows are known to have supersolar metallicities \citep[e.g.][]{Gabel06,Arav07}.
In our Cloudy models the discrepancy persists with different metallicities since the abundance scaling schema from grid M5a of \cite{Hamann93} does not increase the relative [S/P] or [S/Al] abundances. An alternative enrichment model that results in [S/P]=-0.9 and [S/Al]=-0.9 for a certain metallicity would decrease the \siv\ column density predicted by the solution from the \pv\ and \aliii\ constraints by a factor of eight, thus eliminating the discrepancy that appeared in the solar abundances case. Considering the absence of a complete AGN abundance scaling model, this solution is not as implausible as it seems; e.g. \cite{Ballero08} derives relative abundances such as [Si/C]=0.83 for $Z=7.22\Zsun$.

If neither of these scenarios apply, then we conclude that the \siv\ column density is indeed a factor of eight higher than was measured from the \siv\ and \siv* troughs, and one or both of the troughs must be saturated. \siv* cannot be saturated since its trough is shallower than \siv\ (see discussion in Section 6.3 of Paper I), thus the true \siv\ column density increases by a factor of eight, and the \siv*$/$\siv\ column density ratio decreases by the same factor. This correction would decrease \ne\ by an order of magnitude (see \figref{figExFrac}), which would increase $R$ by a factor of three (see \eqnref{eqnUDef}), and increase \mdot\ and \edot\ by a factor of three (see \eqnref{ekeq}). Although this effect would enhance the energetics of the outflow, we suspect that \siv\ is not saturated since the depth of \siv\ $\lambda1062.66$ is shallower than either line from the \pv\ doublet which is presumed much closer to saturation. As \siv\ and \pv\ are both high-ionization ions, we assume that they have similar covering factors and thus their absorption lines would be the same depth if saturated.

\subsection{Comparison with other outflows}\label{secResultsComparison}
In \tabref{tabEnerCompare} we also list four other outflows with the highest \edot\ determined by our group. Two of these are low-ionization BALs that appear in SDSS J0838+2955 \citep{Moe09} and SDSS J0318-0600 \citep{Dunn10a}, and their distance estimate is based on singly ionized species. This introduces an uncertainty (see \secref{secIntro} for discussion) that can be alleviated by utilizing metastable levels from high-ionization ions such as \siv/\siv*.

%In \tabref{tabEnerCompare} we also list four other outflows with the highest \edot\ determined by our group. Two of these are low-ionization BALs that appear in SDSS J0838+2955 \citep{Moe09} and SDSS J0318-0600 \citep{Dunn10a}. In addition to the \civ\ and \siiv\ BALs common to all outflows, these objects also exhibit absorption from singly-ionized species such as \cii, \mgii, \alii, \siii\ and \feii. The analysis of these outflows assumes an $\Omega$ appropriate for \civ\ outflows, but \cite{Dunn10a} (their Section~5.2) argues that $\Omega$ could be lower for such LoBALs. This model dependency is overcome by utilizing ions with an ionization potential similar to \civ, (e.g. \siv). This ensures that any properties determined by \siv/\siv* are applicable to the \civ\ outflow, since both \siv\ and \civ\ are formed in the same region of the outflow.

The first object to show a large kinetic luminosity (\edot=10$^{46}$ erg~s$^{-1}$) using \siv/\siv* absorption was SDSS J1106+1939 (Paper I). The outflow we study in this paper (SDSS J0831+0354) is the second such example, and it exhibits properties similar to SDSS J1106+1939, albeit at one-third the distance to the quasar. In a cursory examination of the objects to be analyzed in our VLT/X-shooter survey, at least two additional BAL outflows show \siv/\siv* troughs.
%The most energetic (\edot=10$^{46}$ erg~s$^{-1}$) of those outflows is found in SDSS J1106+1939 (Paper I) and was the first object to show \siv/\siv* absorption from a high-energy outflow. The outflow we study in this paper (SDSS J0831+0354) is the second such example. In a cursory examination of the objects to be analyzed in our VLT/X-shooter survey, at least two additional BAL outflows show \siv/\siv* troughs.

\cite{Arav13} also used the excited state from a high-ionization ion (\oiv/\oiv*) to determine the energetics of an outflow from quasar HE 0238-1904. Although the outflow did not show absorption from low-ionization ions, absorption from the very-high-ionization ions \neviii\ and \mgx\ confirmed the existence of a high-ionization phase. As discussed in the previous section, the \siv\ and \pv\ absorption in SDSS J0831+0354 could originate from a separate ionization phase, but we cannot confirm its existence without absorption data from additional high-ionization ions. %We notice the resemblence of the high-ionization phase of the HE 0238-1904 outflow (log \uh=0.5) to that of SDSS J0831+0354, which can be inferred from the crossing point of \pv\ and \siv\ (log \uh $>$0.3).

%SDSS J0831+0354 exhibits absorption both from low-ionization species (\mgii\ and \aliii) and from high-ionization species (\pv\ and \siv). From the previous objects, these two categories of ionization species should be separated into two ionization solutions, but additional constraints for both ionization phases are needed to determine the solution (see \secref{secResultsSIVDiscrepancy}). We notice the resemblence of the high-ionization phase of the HE 0238-1904 outflow (log \uh=0.5) to that of SDSS J0831+0354, which can be inferred from the crossing point of \pv\ and \siv\ (log \uh $>$0.3).
%SDSS J0831+0354 exhibits absorption both from low-ionization species and from high-ionization species. Disregarding $N_\mathrm{ion}$ from saturated lines (lower limits in \tabref{tabColDen}) and from non-detected lines (upper limits in \tabref{tabColDen}), measurements from only two low-ionization ions (\mgii\ and \aliii) and two high-ionization ions (\pv\ and \siv) are available for the photoionization analysis. A one-ionization parameter solution is not sufficient to explain these measured column densities from both categories of ionization species, yet those four measurements are not enough to over-constrain a two-ionization phase solution that has four free parameters. Additional constraints for both ionization phases are needed to determine the solution. We notice the resemblence of the high-ionization phase of the HE 0238-1904 outflow (log \uh=0.5) to that of SDSS J0831+0354, which can be inferred from the crossing point of \pv\ and \siv\ (log \uh $>$0.3).

\subsection{Distance of quasar outflows from the central source}\label{secDistance}

The outflow of SDSS J0831+0354 lies closer to the central source than the other outflows listed in \tabref{tabEnerCompare}. However, this distance ($R\sim100$ pc) is three orders of magnitude greater than the trough forming region (0.01--0.1 pc) for accretion disk line-driven winds \citep[e.g. Figure (3) of ][]{Murray95,Proga00}. This large empirical distance scale is not limited to high \edot\ outflows.
The large majority of measured distances to BAL (and narrower) outflows that were deduced using troughs from excited states yield distances between 10--10,000 pc
\citep{deKool01,Hamann01,deKool02b,deKool02c,Moe09,Dunn10a,Bautista10,Aoki11,Edmonds11,Borguet12a,Borguet12b,Borguet13,Arav13,Lucy14}.
Two outflow components are found to be between 1--10 pc from the central source \citep{deKool02b,deKool02c}.
The discrepancy between the empirical distances and those inferred from accretion disk line-driven winds models suggests that 
the latter models are not applicable to the trough formation of quasar outflows.

\subsection{Reliability of measurements}\label{secResultsRely}
Our group \citep{Arav97,Arav99a,Arav99b,Arav01b,Arav01,Arav02,Arav03,Scott04,Gabel05a} and others \citep{Barlow97a,Hamann97b,Telfer98,Churchill99,Ganguly99} showed that column densities derived from the apparent optical depth (AOD) analysis of BAL troughs are unreliable when non-black saturation occurs in the troughs. For this reason, we use partial covering (PC) absorption models in this analysis.

It is instructive to assess quantitatively the difference in \edot\ that arises from using these two absorption models on our results. To this end we re-determine the photoionization solution using the AOD measurements for \pv, \siiii, \aliii\ and \siv\ (i.e. from the first column of \tabref{tabColDen}). The resulting kinetic luminosity of \edot=$10^{45.6}$~erg~s$^{-1}$ for the UV-Soft 4\Zsun\ model differs only slightly from the \edot=$10^{45.7}$~erg~s$^{-1}$ that we derive using the PC template. Thus, our results are rather insensitive to the method of column density extraction. 

In \tabref{tabRobust} we demonstrate that almost all possible deviations from our representative model (UV-Soft SED, $Z=4\Zsun$, one ionization component, unsaturated \siiii\ absorption) results in a higher \edot\ for the SDSS J0831+0354 outflow. Even with \edot\ lowered by assuming a different SED, the minimum energy allowed by our analysis is \edot$/L_\mathrm{Edd}=\revise{2.6\%}$ for the HE0238 SED with $Z=4\Zsun$ (see \tabref{tabEnerCompare}).
%In \tabref{tabRobust} we demonstrate that almost all possible deviations from our representative model (UV-Soft SED, $Z=4\Zsun$, one ionization component, unsaturated \siiii\ absorption) results in a higher \edot\ for the SDSS J0831+0354 outflow. The exception to this is variation of the SED, but even after lowering \edot\ in this way, all else equal, the minimum energy allowed by our analysis is \edot$/L_\mathrm{Edd}=1.2\%$ for the HE0238 SED with $Z=4\Zsun$ (see \tabref{tabEnerCompare}).
%In \tabref{tabRobust} we show how changes in various assumptions affect the deduced \edot. With the exception of varying the SED, any plausible change from our representative model (UV-Soft SED, $Z=4\Zsun$, one ionization component, unsaturated \siiii\ absorption) results in a higher \edot\ for the SDSS J0831+0354 outflow. Even with \edot\ lowered by assuming a different SED, the minimum energy allowed by our analysis is \edot$/L_\mathrm{Edd}=1.2\%$ for the HE0238 SED with $Z=4\Zsun$ (see \tabref{tabEnerCompare}).

\begin{table}
\caption{Sensitivity of derived energetics to input changes.}
\label{tabRobust}\begin{tabular}{lcrl}\hline
Assumption			&Effect(s)		&Outcome		&Reference\\\hline
N(\siiii)$>N_\mathrm{AOD}$	&higher \Nh,		&higher \edot		&\secref{secPhotoSoln}\\
				&lower \uh		&			&\\
$Z<4\Zsun$			&higher \Nh		&higher \edot		&\figref{figMultiModel}\\
different SED			&lower \Nh		&lower \edot		&\figref{figMultiModel}\\
$\Omega>0.08$			&---			&higher \edot		&\eqnref{ekeq}\\
two phases			&higher \Nh,		&higher \edot		&\secref{secResultsSIVDiscrepancy}\\
				&higher \uh		&			&\\
\siv* saturated			&lower \nh,		&higher \edot		&\secref{secResultsSIVDiscrepancy}\\
				&higher $R$		&			&\\\hline
\end{tabular}\vspace{-0.7\baselineskip}\end{table}

\section{SUMMARY}\label{secSummary}
We present the analysis of BALQSO SDSS J0831+0354 using data obtained with VLT/X-shooter. Our main findings are as follows:
\newcounter{summary}\begin{list}{\arabic{summary}.}{\usecounter{summary}\setlength{\itemindent}{0cm}\setlength{\labelwidth}{0.65\parindent}\setlength{\leftmargin}{\labelwidth}}
\item{The BAL outflow in SDSS J0831+0354 has a velocity of \revise{10,800 \kms} in the rest frame of the quasar with a width of 3500 \kms\ and shows kinematically connected absorption from the high-ionization species \siv\ and \siv*. The ratio of the excited column density to ground was used to determine the number density of the outflow to be log (\ne)=4.42$\pm$0.24 cm$^{-3}$.}
%\item{The non-black saturation of the \pv\ doublet indicated that the absorber was inhomogenously distributed across the emission source. Photoionization models that predict such large absorption from \pv\ neccessitate a high column density (\Nh) and high-ionization (\uh) outflow.}
\item{The non-black saturation of the \pv\ doublet indicated that the absorber was inhomogenously distributed across the emission source. The saturation of the \pv\ doublet resulted in a high ionic column density which, combined with the low elemental abundance of phosphorous, necessitates a high total hydrogen column density (\Nh).}
%The saturation of the \pv\ doublet resulted in a high ionic column density which, combined with the low elemental abundance of phosphorous, neccessitates a high column density (\Nh) and high-ionization outflow.
\item{Variations of the incident SED had only a small effect on the ionization solution (\uh,\Nh) of the outflow, while an increased metallicty of 4\Zsun\ lowered the required hydrogen column density by 0.4 dex. The UV-soft SED with Z=4\Zsun\ was adopted as the representative model, and the hydrogen column density was determined to be log~(\Nh)=22.0$\pm$0.4~cm$^{-2}$ with an ionization parameter of log~(\uh)=-0.25$\pm$0.3.}
%\item{The photoionization solution was determined from two high-ionization (\pv\ and \siv) and two low-ionization (\mgii\ and \aliii) ions. The solution over-predicts the observed \siv\ column density, but this discrepancy can be resolved either with a two-ionization phase solution or with individual elemental abundance scaling.}
\item{The outflow was determined to be located \revise{110 pc} from the central source and contained a kinetic luminosity of \revise{5.2\%} of the Eddington luminosity of the quasar. We establish half of this value (\revise{2.6\%}) as a conservative lower limit. This large kinetic luminosity makes the outflow of SDSS J0831+0354 a candidate for quasar-mode AGN feedback effects seen in simulations \citep[e.g.][]{Hopkins06,Ostriker10}.}
\end{list}

\section*{ACKNOWLEDGMENTS}
We acknowledge support from NASA STScI grants GO 11686 and GO 12022 as well as NSF grant AST 1413319.

%\begin{figure}\includegraphics[width=\columnwidth]{Outflows-in-NGC5548_4-28-14_FINAL_500k.eps}
%\caption{NGC5548 Concept image test.}
%\label{figNGC5548Concept}\end{figure}

\bibliographystyle{mn2e}
\bibliography{astro}{}
\end{document}